\documentclass[prb,aps,twocolumn,amstex,preprintnumbers,amsmath,amssymb,array,tabularx]{revtex4}
\usepackage{graphicx}
\usepackage{dcolumn}
\usepackage{bm}
\usepackage{lscape} 
\usepackage{amsmath}
\usepackage{amsxtra}
\usepackage{tabularx}
\usepackage{setspace}

\DeclareGraphicsExtensions{.jpg,.pdf,.eps,.gif}

\begin{document}
\title{Half-metallic zigzag carbon nanotube dots}

\author{$\mbox{Oded Hod}$ and $\mbox{Gustavo E. Scuseria}$}

\affiliation{Department of Chemistry, Rice University, Houston,
  Texas 77005-1892}

\date{\today}

\begin{abstract}
  A comprehensive {\it first-principles} theoretical study of the
electronic properties and half-metallic nature of finite zigzag carbon
nanotubes is presented. Unlike previous reports, we find that all
nanotubes studied present a spin-polarized ground state, where
opposite spins are localized at the two zigzag edges in a long-range
antifferomagnetic configuration.  Relative stability analysis of the
different spin states indicate that, for the shorter segments,
spin-ordering should be present even at room temperature. The energy
gap between the highest occupied and the lowest unoccupied molecular
orbitals of the finite systems is found to be inversely proportional
to the nanotubes segments length, suggesting a route to control their
electronic properties.  Similar to the case of zigzag graphene
nanoribbons, half-metallic behavior is obtained under the influence of
an external axial electric field.

\end{abstract}

\maketitle
 


\section{Introduction}
Magnetism in carbon based materials has attracted considerable
scientific interest in recent years both from an
experimental~\cite{Shibayama2000, Prasad2000, Kopelevich2000,
  Makarova2001, Takai2001, Coey2002, Esquinazi2002, Wood2002,
  Hohne2002, Esquinazi2003, Makarova2004, Esquinazi2005, Mombru2005,
  Makarova2006, Ohldag2007, Tombros2007, Likodimos2007} and a
theoretical~\cite{Kobayashi1993, Klein1994, Fujita1996, Nakada1996,
  Wakabayashi1998, Nakada1998, Wakabayashi1999, Miyamoto1999,
  Kawai2000, Okada2001, Kusakabe2003, Yamashiro2003, Lehtinen2003,
  Andriotis2003, Okada2003, Kim2003, Park2003, Hikihara2003, Ma2004,
  Higuchi2004, Kusakabe2004, Ryu2004, Lehtinen2004-1, Lehtinen2004-2,
  Lee2005, Ma2005, Chen2005, Son2006-1, Son2006-2, Hod2007-1,
  Hod2007-2, Rossier2007, Ezawa2007, Ezawa2008, Hod2008, Pisani2008}
viewpoint.  While the origin of magnetic ordering in such systems is
yet to be fully understood, it has been suggested that spin
polarization may arise from local structural defects,~\cite{Kim2003,
  Ma2004, Lehtinen2004-2, Pisani2008} sterically protected carbon
radicals~\cite{Park2003} and chemical
impurities.~\cite{Lehtinen2004-1, Lehtinen2003, Ma2005} A unique
mechanism for spin ordering in graphene based systems is related to
the appearance of edge states.~\cite{Kobayashi1993, Klein1994,
  Fujita1996, Nakada1996, Wakabayashi1998, Nakada1998,
  Wakabayashi1999, Miyamoto1999, Kawai2000, Okada2001, Kusakabe2003,
  Yamashiro2003, Lee2005, Son2006-1, Son2006-2, Hod2007-1, Hod2007-2,
  Rossier2007, Ezawa2007, Ezawa2008, Hod2008} When cutting a graphene
sheet along its zigzag axis to form a narrow and elongated graphene
nanoribbon (GNR), distinct electronic states appear, which are
localized around the exposed edges.~\cite{Giunta2001, Klusek2005,
  Kobayashi2005, Niimi2005, Niimi2006, Kobayashi2006} These states are
predicted to carry spin polarization, resulting in a well defined
magnetic ordering.~\cite{Klein1994, Fujita1996, Nakada1996,
  Wakabayashi1998, Nakada1998, Wakabayashi1999, Miyamoto1999,
  Kawai2000, Kusakabe2003, Yamashiro2003, Lee2005, Son2006-1,
  Son2006-2, Hod2007-1, Hod2007-2, Rossier2007, Ezawa2007, Ezawa2008,
  Hod2008} Due to the bipartite hexagonal structure of graphene, the
electronic ground state of such zigzag graphene nanoribbons (ZZ-GNRs)
is characterized by an antiferromagnetic (AFM) spin ordering, where
the edge states located at the two zigzag edges of the ribbon have
opposite spins.  Under the influence of an external electric field,
these systems have been predicted to become half-metallic with one
spin channel being semiconducting and the other
metallic,~\cite{Son2006-1, Hod2007-1} thus acting as perfect spin
filters with important implications to the field of spintronics.
Furthermore, this behavior has been shown to be preserved for finite
graphene nanoribbons, where the zigzag edge may be as short as a few
repeating units.~\cite{Hod2007-2, Rossier2007, Ezawa2007, Ezawa2008,
  Hod2008}

Similar to graphene nanoribbons, is has recently been suggested that
spin ordering may occur on the zigzag edges of hydrogen terminated
finite sized carbon nanotubes (CNTs).~\cite{Okada2003, Higuchi2004}
Based on density functional theory (DFT) calculations within the local
spin density approximation (LSDA), it was predicted that the ground
electronic state of these systems strongly depends on their
circumferencial dimension ranging from AFM to ferromagnetic (FM)
ordering, where both edges of the nanotube bear spins of the same
flavor. This conclusion is in striking difference from the case of
ZZ-GNRs, where the ground electronic state was found to have an AFM
spin ordering regardless of the dimensions of the
ribbon.~\cite{Klein1994, Fujita1996, Nakada1996, Wakabayashi1998,
  Nakada1998, Wakabayashi1999, Miyamoto1999, Kawai2000, Kusakabe2003,
  Yamashiro2003, Lee2005, Son2006-1, Son2006-2, Hod2007-1, Hod2007-2,
  Rossier2007, Ezawa2007, Ezawa2008, Hod2008}

Interestingly, contradicting results have been reported in the
literature,~\cite{Kim2003} indicating that the ground state of
unpassivated zigzag CNTs has a low-spin AFM ordering even for
nanotubes that have earlier been considered to present a high-spin
ferromagnetically ordered ground state.~\cite{Okada2003}

In light of the considerable progress that has been made in the
synthesis and fabrication of ultra-short~\cite{Gu2002, Nakamura2003,
  Javey2004, Chen2006-1, Chen2006-2} and open ended~\cite{Ajayan1993}
CNTs and ultra-narrow GNRs,~\cite{Han2007, Chen2007, Li2008, Wang2008}
it is desirable to obtain a full understanding of their electronic
properties.  Such understanding may prove important in future
applications of these molecules in nanoelectronic and nanospintronic
devices.

\input{epsf}
\begin{figure}[h]
\begin{center}
\epsfxsize=8.65cm \epsffile{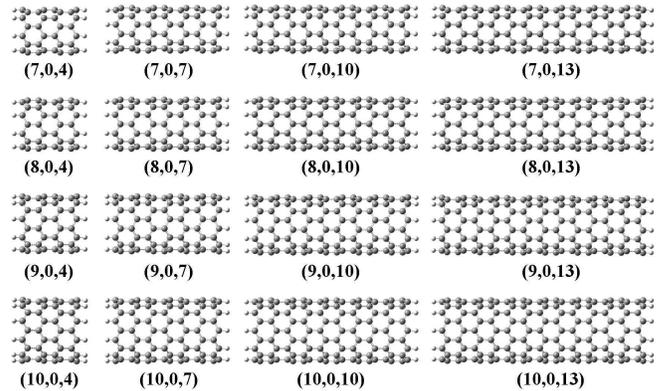}
\end{center}
\caption{The four sets of CNTs studied. The notation $(N,M,L)$
  indicates a finite segment of a $(N,M)$ CNT with $L$ zigzag rings
  along its principal axis.}
\label{Fig: CNTs}
\end{figure}

It is the purpose of the present paper to present a comprehensive
investigation of spin polarization in hydrogen passivated finite
zigzag CNTs.  Unlike previous reports on these
systems,~\cite{Okada2003, Higuchi2004} we find that all zigzag CNTs
studied present an antiferromagnetically ordered ground state,
regardless of their diameter and length.  Furthermore, as is the case
with infinitely long~\cite{Son2006-1, Son2006-2, Hod2007-1} and
finite~\cite{Hod2008} GNRs, the application of an external axial
electric field results in half-metallic behavior.

\section{Method}
We have studied a set of sixteen finite segments of the (7,0), (8,0),
(9,0), and (10,0) zigzag CNTs (see Fig.~\ref{Fig: CNTs}). For each CNT
four segments of 0.93, 1.56, 2.20, and 2.84 nm in length are
considered.  All nanotubes considered are hydrogen terminated, i.e.,
each carbon edge atom is passivated with a single hydrogen atom.  We
label the CNTs by $(N,M,L)$, where $(N,M)$ is the infinite CNT from
which the relevant finite segment is derived and $L$ is the number of
zigzag carbon rings stacked together to form the finite CNT (see
Fig.~\ref{Fig: CNTs}).  This set of nanotubes includes all those
previously predicted to present a ferromagnetically ordered ground
state.~\cite{Okada2003}

All the calculations presented in this work were carried out using the
development version of the {\it Gaussian} suite of
programs.~\cite{Frisch2004} Spin polarized ground state calculations
were performed using the screened exchange hybrid density functional
of Heyd, Scuseria and Ernzerhof (HSE06),~\cite{Heyd2003, Heyd2006,
  Izmaylov2006, Keywords} which has been tested on a wide variety of
materials and has been shown to accurately reproduce experimental
bandgaps~\cite{Heyd2004, Heyd2005, Brothers2008} and first and second
optical excitation energies in metallic and semiconducting single
walled CNTs.~\cite{Barone2005-1, Barone2005-2} The inclusion of
short-range exact-exchange in the HSE06 functional makes it suitable
to treat electronic localization effects,~\cite{Kudin2002, Prodan2005,
  Prodan2006, Hay2006, Kasinathan2006} which are known to be important
in this type of materials.~\cite{Kobayashi1993, Klein1994, Fujita1996,
  Nakada1996, Wakabayashi1998, Nakada1998, Wakabayashi1999,
  Miyamoto1999, Kawai2000, Okada2001, Kusakabe2003, Yamashiro2003,
  Niimi2005, Kobayashi2005, Lee2005, Son2006-1, Son2006-2, Niimi2006,
  Kobayashi2006,Hod2007-1, Hod2007-2, Hod2008} This is further
supported by the good agreement, which was recently obtained, between
predicted bandgaps~\cite{Barone2006} of narrow graphene nanoribbons
and measured values.~\cite{Han2007, Chen2007, Li2008} To obtain a
reliable ordering of the different magnetization states, we find it
important to relax the geometry of the finite CNTs for each spin
polarization.  Therefore, unless otherwise stated, all reported
electronic properties are given for fully optimized structures using
the double-zeta polarized 6-31G** Gaussian basis
set.~\cite{Hariharan1973} It should be noted that since our
calculations are performed within a single determinantal framework, we
can determine only the total spin vector projection along a given
axis, $m_s$, and not the total spin. This is standard in unrestricted
Kohn-Sham theory.

\section{Results and discussion}
\begin{figure}[h]
\begin{center}
\epsfxsize=8.65cm \epsffile{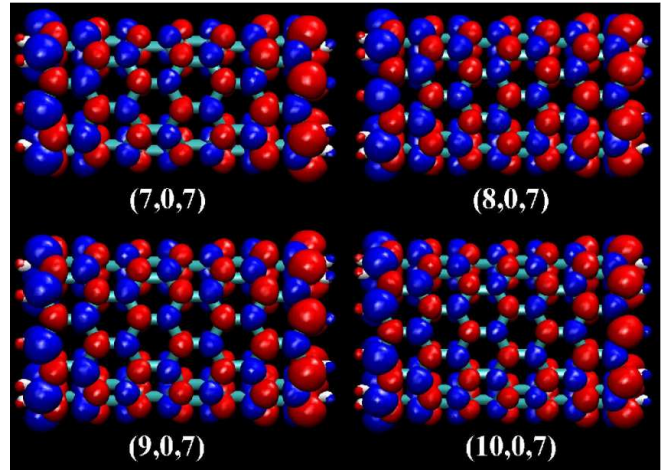}
\end{center}
\caption{Antiferromagnetic ground state spin density maps of the
  (7,0,7) (upper left panel), (8,0,7) (upper right panel), (9,0,7)
  (lower left panel)), (10,0,7) (lower right panel), finite zigzag
  CNTs as obtained using the HSE06 functional with the 6-31G** basis
  set.  Red and blue isosurfaces indicate the two spin flavors with an
  isovalue of 0.0015 $a_0^{-3}$.}
\label{Fig: Spin Density}
\end{figure}

We start by performing ground state calculations on the full set of
sixteen finite CNTs studied.  In Fig.~\ref{Fig: Spin Density}
spin-density maps of the ground state of four representative CNT
segments are presented.  In contradiction with previously reported
LSDA results,~\cite{Okada2003, Higuchi2004} which predict the $(8,0)$
CNT derivatives to have an AFM ground state and the $(7,0)$ and
$(10,0)$ nanotube segments to have a ferromagnetically ordered ground
state, our HSE06 calculations indicate that all CNTs studied have an
antiferromagnetically ordered ground state, with a total spin vector
projection of $m_s=0$ and a spatially resolved spin density.  Similar
to the case of finite GNRs,~\cite{Hod2007-2, Hod2008} one zigzag edge
of the CNTs segments has a high density of spin $\alpha$ electrons
(blue color in Fig.~\ref{Fig: Spin Density}) while the other edge is
rich with spin $\beta$ electrons (red color in the figure).

\input{epsf}

To understand the origin of this discrepancy between the predictions
obtained using the semi-local and the screened-hybrid functionals, we
have repeated the calculations using the LSDA functional and the same
atomic basis set. Interestingly, we encountered considerable
convergence problems when performing the calculations at this level of
theory.  For many of the systems we have found it necessary to start
the SCF with a pre-designed initial guess in order to achieve
convergence to the appropriate spin state.  Once convergence was
achieved we have found that for the $(7,0,7)$ and $(7,0,10)$ systems
the LSDA functional does predict a ferromagnetically ordered ground
state.  Nevertheless, this state is only $5$ meV below the
antiferromagnetically ordered state.  For the rest of the systems
considered, whenever convergence was achieved, an AFM ground state has
been found.

We attribute this behavior to the self-interaction error appearing in
the LSDA functional, which tends to considerably over estimate
electron delocalization.~\cite{Ruzsinszky2006, Sanchez2006} Since some
of the most important physical characteristics of the systems under
consideration relate to pronounced localized edge states, care should
be taken when using the LSDA functional to capture the correct
energetic ordering of the different spin polarized states.  As
mentioned above, the inclusion of Hartree-Fock exchange within hybrid
functionals considerably reduces the delocalization error and thus
makes them more appropriate to treat scenarios where electron
localization plays an important rule in determining the electronic
structure of the system.~\cite{Kudin2002, Prodan2005, Prodan2006,
  Hay2006, Kasinathan2006}

\input{epsf}
\begin{figure}[h]
\begin{center}
\epsfxsize=8.65cm \epsffile{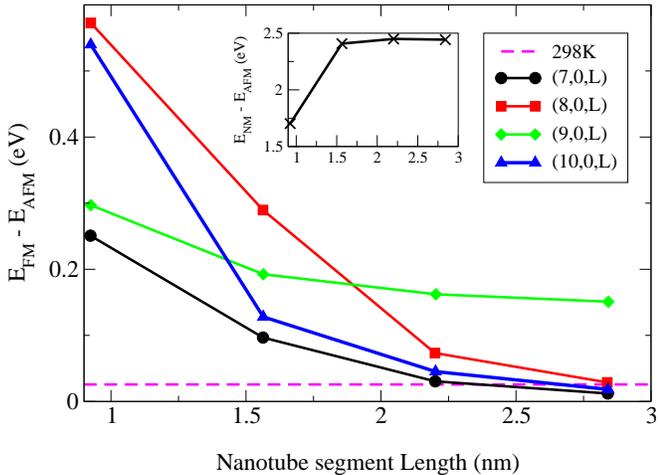}
\end{center}
\caption{Energy differences between the antiferromagnetic ground state
  and the above lying higher spin multiplicity ferromagnetic state for
  the four sets of CNTs segments studied, as calculated by the HSE06
  functional.  $k_B$T at room temperature is indicated by the dashed
  magneta line.  Inset: Energy difference between the
  antiferromagnetic ground state and the non-magnetic closed shell
  electronic state of the $(10,0,L)$ CNT segments.}
\label{Fig: Stability}
\end{figure}

To quantify our findings, we study the stability of the
antiferromagnetically ordered ground state with respect to the above
lying ferromagnetically ordered spin state.  In Fig.~\ref{Fig:
  Stability} the energy differences between the antiferromagnetic
ground state and the first higher spin multiplicity state are
presented for the four sets of nanotube segments studied.  We indicate
$k_B$T ($k_B$ being Boltzmann's constant) at room temperature (T=298
K) by the dashed magenta (color online) line in the figure.  For the
$(7,0,L)$ and $(9,0,L)$ systems we find the $m_s=2$ state to be the
above lying ferromagnetic state.  For the $(10,0,L)$ segments the
first higher spin multiplicity state has $m_s=3$.  In the case of the
$(8,0,L)$ segments the higher spin multiplicity state changes from
$m_s=1$ for $L=4$ to $m_s=3$ for $L=7, 10$ and $13$. As can be seen,
the AFM ground state of the shorter CNT segments studied is
considerably more stable than the above lying higher spin state, and
is expected to be detectable at room temperature.  Like in the case of
GNRs,~\cite{Son2006-1} the energy difference between the ground state
and the above lying higher spin-state decreases monotonically with
increasing CNT length, up to a point where the differences become
lower than room temperature.  Interestingly, we find that for the
$(9,0)$ segments, which in the limit of infinite length become
metallic, the decay rate of the energy difference between the AFM
ground state and the above lying FM state is slower than that of the
other semiconducting tubes studied.  Even for this system an
exponential fitting of the decay curve suggests that at a length of
$8$ nm the energy difference is below room temperature.  The inset of
Fig.~\ref{Fig: Stability} shows the energy difference between the AFM
ground state and the non-magnetic closed-shell electronic state of the
$(10,0,L)$ segments.  The non-magnetic state is found to be
considerably higher in energy when compared to higher spin
multiplicity states.  This, as well, is in agreement with the case of
GNRs~\cite{Son2006-1, Hod2008} and suggests that even at room
temperature short segments of CNTs are expected to present magnetic
properties.

\input{epsf}
\begin{figure}[h]
\begin{center}
\epsfxsize=8.65cm \epsffile{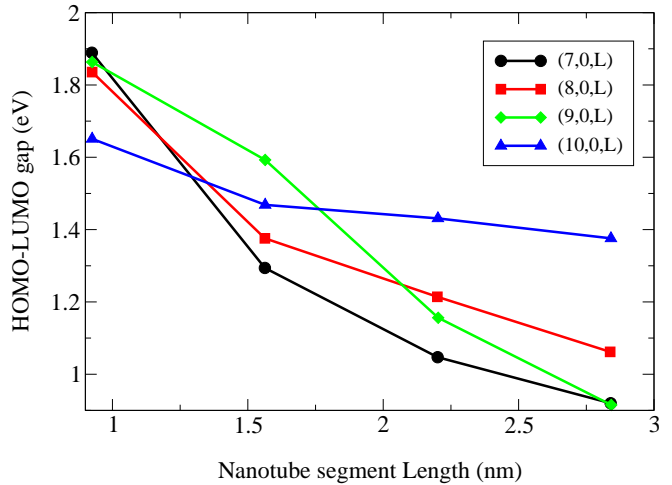}
\end{center}
\caption{HOMO-LUMO gap values as a function of CNT segment length for
  the four sets of CNTs studied as calculated by the HSE06
  functional.}
\label{Fig: Gaps}
\end{figure}

Before discussing the effect of an electric field on the electronic
properties of finite CNTs, it is essential to study their ground state
characteristics in the absence of external perturbations.  The length
dependence of the HOMO-LUMO (highest occupied molecular orbital and
lowest unoccupied molecular orbital, respectively) gap would be the
most important parameter to address.  In Fig.~\ref{Fig: Gaps} the
energy gap as a function of the length of the CNTs segments are
presented, for the four subsets of nanotubes considered.  All studied
finite zigzag CNT segments have sizable HOMO-LUMO gaps, including the
$(9,0,L)$ series, which in the limit of $L\to\infty$ is expected to
become metallic.  The HOMO-LUMO gap is inversely proportional to the
length of the CNTs segments for all nanotubes studied and its value
strongly depends on the length, changing by more than $0.8$ eV for
lengths in the range of $0.9$ and $2.8$ nm.  This suggest that careful
tailoring of the nanotube length can be used as a sensitive control
parameter over its electronic properties.~\cite{Rochefort1999}

\input{epsf}
\begin{figure}[h]
\begin{center}
\epsfxsize=8.65cm \epsffile{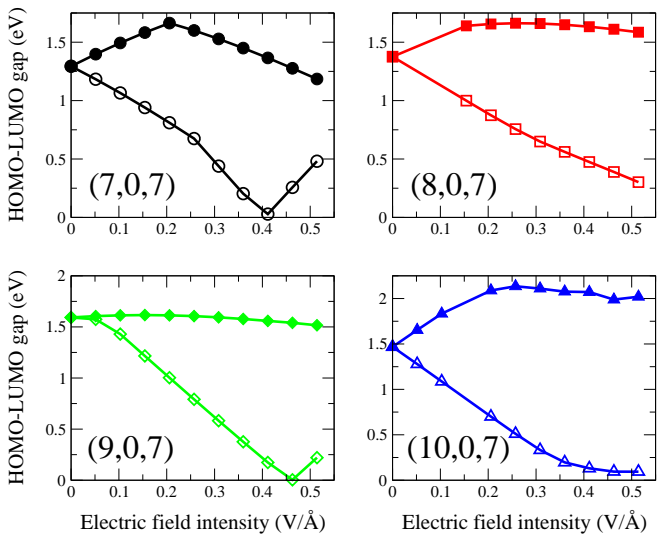}
\end{center}
\caption{Spin-polarized HOMO-LUMO gap dependence on the strength of an
  external axial electric field for the $(7,0,7)$ (upper left panel),
  $(8,0,7)$ (upper right panel), $(9,0,7)$ (lower left panel), and
  $(10,0,7)$ (lower right panel) finite zigzag CNTs as calculated by
  the HSE06 functional.  Fixed geometries of the relaxed structures in
  the absence of the external field were used.  Full and open marks
  stand for the $\alpha$ and $\beta$ spin gaps, respectively.}
\label{Fig: EField}
\end{figure}

We now turn to check whether the similarity between finite zigzag CNTs
and finite GNRs extends also to the case of half-metallic behavior
under the influence of an external electric field.~\cite{Son2006-1,
  Hod2008} In Fig.~\ref{Fig: EField} we present the spin-polarized
HOMO-LUMO gap dependence on an external electric field applied
parallel to the main axis of the nanotubes (perpendicular to the
zigzag edges) for four representative finite zigzag CNTs.  In
agreement with previous calculations,~\cite{Son2006-1, Hod2007-1,
  Hod2008} the $\alpha$ and $\beta$ gaps are degenerate in the absence
of an external field.  Upon application of the field, electrons having
one spin experience an increase in the HOMO-LUMO gap while the
opposite spin flavor experiences a decrease in the gap.  This gap
splitting continues up to a point, where one spin channel presents a
vanishing gap, thus creating a half-metallic state or a perfect spin
filter. At this point, due to the increased mobility of the metallic
electrons, further increase in the external field results in spin
transfer between both edges thus reducing the total spin polarization
and the energy gap splitting (clearly seen on the upper left panel of
Fig.~\ref{Fig: EField}).  All four representative structures present
the same features described above.  The main differences observed are
in the zero field HOMO-LUMO gap and the onset field for the appearance
of the half-metallic state.  Nevertheless, they all predict a
half-metallic state at an appropriate electric field strength.

\section{Conclusions}
To summarize, intrigued by reports of contradicting results regarding
the ground state properties of finite zigzag carbon nanotubes, we
performed a comprehensive study of the electronic character of a
representative set of sixteen CNT segments.  Unlike previously
reported results, we found that all finite zigzag CNTs possess a
spin-polarized ground state with antiferromagnetic spin ordering.
This state is characterized by high spin-density of opposite spins
located at the two zigzag edges of the molecule. Our HSE06 results
predict the antiferromagnetic ordering to be considerably stable with
respect to higher spin multiplicity states and the non-magnetic closed
shell state, suggesting that their spin polarization is detectable at
room temperature.  The HOMO-LUMO gap was found to be inversely
proportional to the length of the zigzag CNT segment.  The high
sensitivity of the gap to changes in the length of the CNT segment
suggests a way to control the electronic properties of such systems.
Similar to the case of zigzag GNRs, the half-metallic nature of finite
zigzag CNTs under an external in-plane electric field was verified.
Due to the recent success of the HSE06 functional in predicting the
electronic properties of GNRs and CNTs of different nature and
dimensions, we are confident about the reliability of the predictions
presented here.



\section{Acknowledgments}
This work was supported by NSF CHE-0807194 and the Welch Foundation
(C-0036). Calculations were performed in part on the Rice Terascale
Cluster funded by NSF under Grant EIA-0216467, Intel, and HP.  O.H.
would like to thank the generous financial support of the Rothschild
and Fulbright foundations.


\bibliographystyle{./prsty.bst} \bibliography{Finite-CNT}
\end{document}